# Long distance transport of magnon spin information in a magnetic insulator at room temperature

L.J. Cornelissen[1]*, J.Liu[1], R.A. Duine[2], J. Ben Youssef[3] & B. J. van Wees[1]

[1]*Zernike Institute for Advanced Materials, Physics of Nanodevices, University of Groningen, 9747 AG Groningen, The Netherlands*

[2]*Institute for Theoretical Physics and Center for Extreme Matter and Emergent Phenomena, University of Utrecht, 3512 JE Utrecht, The Netherlands*

[3]*Université de Bretagne Occidentale, Laboratoire de Magnétisme de Bretagne CNRS, 6 Avenue Le Gorgeu, 29285 Brest, France*

* L.J.Cornelissen@rug.nl

**The transport of spin information has been studied in various materials, such as metals[1], semiconductors[2] and graphene[3]. In these materials, spin is transported by diffusion of conduction electrons[4]. Here we study the diffusion and relaxation of spin in a magnetic insulator, where the large bandgap prohibits the motion of electrons. Spin can still be transported, however, through the diffusion of non-equilibrium magnons, the quanta of spin wave excitations in magnetically ordered materials. Here we show experimentally that these magnons can be excited and detected fully electrically[5,6] in linear response, and can transport spin angular momentum through the magnetic insulator yttrium iron garnet (YIG) over distances as large as 40 μm. We identify two transport regimes: the diffusion limited regime for distances shorter than the magnon relaxation length, and the relaxation limited regime for larger distances. With a model similar to the diffusion-relaxation model for electron spin transport in (semi)conducting materials, we extract the magnon relaxation length $\lambda = 9.4 \pm 0.6$ μm in a 200 nm thin YIG film at room temperature.**

Recently, a great deal of attention is devoted to the investigation of thermally excited magnons, particularly in studies of the spin Seebeck effect in YIG[7–11], for which the relaxation length of magnons was investigated in a local longitudinal spin Seebeck geometry[12]. Long distance transport of thermally excited magnons was demonstrated very recently at $T = 23$ K, using a non-local geometry[13]. However, thermal excitation is a non-linear and relatively slow process and does not allow for high fidelity transport and conversion of information. To facilitate magnonic devices operating in linear response at room temperature, the ideal signal pathway would be: input electronic charge signal → electron spins → magnons → electron spins → output charge signal. Information processing and transport can then be done in the magnon part of the pathway.

Kajiwara et al.[14] reported long distance transmission of signals in YIG by spin waves interconversion. However, in their experiment spin waves are excited by exerting a spin transfer torque[15,16] (STT) on the magnetization large enough to overcome the intrinsic and interfacial Gilbert damping, driving the sample into ferromagnetic resonance (FMR). The STT was generated by the spin Hall effect (SHE) in a platinum layer, deposited on the YIG. Spin waves generated in this type of experiment typically have frequencies $f < 10$ GHz[14,17,18], much smaller than the thermal energy ($hf \ll k_B T$) and are hence in the classical regime. Excitation of this type of spin waves by STT is a highly non-linear process, where a threshold current density has to be overcome to compensate the damping of the specific spin wave modes. Their experiments proved difficult to reproduce, but recently Collet et al. have shown that YIG nanostructures can be driven into FMR via STT[19]. Chumak et al.[20] demonstrated long range low frequency spin wave spin transport using radio frequency magnetic fields to excite spin waves, which were detected making use of the inverse spin Hall effect (ISHE) in platinum.

Here, we demonstrate for the first time the excitation and detection of high frequency magnons (i.e. quantized spin waves) via a spin accumulation in a paramagnetic normal metal layer, and their long distance transport in YIG. Since the spin accumulation can be induced (via the SHE) and detected (via the ISHE) electrically, this method allows for full electrical excitation and detection of magnon spin signals in linear response and provides a new route towards the development of low-power electronic devices, utilizing magnons rather than electrons for the transport and processing of information.

We study the transport of magnons in a non-local geometry, shown schematically in figure 1a. The devices consist of platinum (Pt) strips, deposited on a thin YIG film (see Methods for fabrication details). One Pt strip functions as injector, another as detector (figure 1c). When a charge current $I$ is sent through the injector, the SHE generates a transverse spin current. A spin accumulation $\boldsymbol{\mu_s}$ then builds up at the Pt|YIG interface, pointing in the film plane. When the spin orientation of $\boldsymbol{\mu_s}$ is parallel (antiparallel) to the average magnetization $\boldsymbol{M}$, magnons are annihilated (excited), resulting in a non-equilibrium magnon population $n_m$ in the YIG[5,6] (shown schematically in figure 1a and 1b). The non-equilibrium magnons diffuse in the YIG, giving a magnon current $j_m$ from injector to detector. At the detector, the reciprocal process occurs: magnons interact at the interface, flipping the spins of electrons and creating a spin imbalance in the platinum (figure 1b). Due to the ISHE, the induced spin current is converted into charge current, which under open circuit conditions generates a voltage $V$. The non-local resistance is then $R_{nl} = \frac{V}{I}$.

Since only the component of $\boldsymbol{\mu_s}$ collinear to $\boldsymbol{M}$ contributes to magnon injection/detection, we expect to see a dependence of $R_{nl}$ on the angle $\alpha$ (figure 1c) between the sample and an in-plane external magnetic field $\boldsymbol{B}$ that orients $\boldsymbol{M}$ (see Methods). We perform non-local measurements as a function of $\alpha$ by rotating the sample in a fixed external field. Using lock-in amplifiers, we separate higher order contributions in the voltage by measuring higher harmonics, using: $V = R_1 I + R_2 I^2 + \ldots$, where $R_i$ is the $i$-th harmonic response[21]. Since magnon spin injection/detection scales linear with $I$, its magnitude is obtained from the first harmonic signal. Any thermal effects due to Joule heating (for which $\Delta T \propto I^2$) will be detected in the second harmonic signal. The result of such a measurement is shown in figure 2a (2c) for the first (second) harmonic, and the observed angular dependence is explained schematically in figure 2b (2d).

We fabricated two series of devices with different injector-detector separation distances $d$. Series A is tailored to the short distance regime ($d < 5$ μm), while series B explores the long distance regime ($3 < d < 50$ μm). For each device, a non-local measurement as shown in figure 2 was performed. The magnitudes of the non-local resistances were extracted for every $d$, by fitting the data with:

$$R^{1\omega} = R_0^{1\omega} + R_{nl}^{1\omega} \cos^2(\alpha), \quad (1)$$

$$R^{2\omega} = R_0^{2\omega} + \frac{1}{2} R_{nl}^{2\omega} \cos(\alpha), \quad (2)$$

where $R_0^{1\omega}$ and $R_0^{2\omega}$ are offset resistances (see Methods) and $R_{nl}^{1\omega}$ ($R_{nl}^{2\omega}$) are the magnitudes of the first (second) harmonic signal. Figure 3a and 3c (3b and 3d) show the results on a linear and logarithmic scale, for the first (second) harmonic non-local resistance. Both $R_{nl}^{1\omega}$ and $R_{nl}^{2\omega}$ are normalized to device length, to compare devices having different lengths.

From figure 3 we can clearly observe two regimes, which we interpret as follows: At large distances, signal decay is dominated by magnon relaxation and is characterized by exponential decay. For distances shorter than the magnon relaxation length we observe diffusive transport and the signal follows a $1/d$ behavior (inset figure 3a). Both regimes are well described with a single model, using the 1D spin diffusion equation[22], adapted for magnon transport:

$$\frac{d^2 n_m}{dx^2} = \frac{n_m}{\lambda^2}, \text{ with } \lambda = \sqrt{D\tau}, \quad (3)$$

where $n_m$ is the non-equilibrium magnon density, $\lambda$ is the magnon relaxation length in YIG, $D$ is the magnon diffusion constant and $\tau$ the magnon relaxation time. The 1D approach is valid since the YIG thickness (200 nm) is much smaller than the injector-detector distance $d$, while the device length is much larger than $d$. We assume strong spin-magnon coupling between YIG and platinum, given the large spin-mixing conductance at the Pt|YIG interface[23,24] and the strong spin-orbit interaction in platinum. We therefore impose the boundary conditions $n_m(0) = n_0$ and $n_m(d) = 0$, where $n_0$ is the injected magnon density which is proportional to the applied current and is determined by various material and interface parameters. These conditions imply that the injector acts as a low impedance magnon source, and all magnon current is absorbed when it arrives at the detector. The solution to equation 3 is of the form $n_m(x) = a\exp(-x/\lambda) + b\exp(x/\lambda)$, and from the boundary conditions we find for the magnon diffusion current density $j_m = -D\frac{dn_m}{dx}$ at the detector:

$$j_m(x = d) = -2D\frac{n_0}{\lambda}\frac{\exp(d/\lambda)}{1-\exp(2d/\lambda)}. \qquad (6)$$

The non-local resistance is proportional to $\frac{j_m(d)}{n_0}$, and we adopt a two-parameter fitting function for the non-local resistances, capturing the distance independent prefactors in a single parameter $C$:

$$R_{nl} = \frac{C}{\lambda}\frac{\exp(d/\lambda)}{1-\exp(2d/\lambda)}. \qquad (7)$$

The signal decay described by equation 7 is equivalent to that of spin signals in metallic spin valves with transparent contacts[25]. The dashed lines shown in figure 3a,c are best fits to this function, where we find from the first harmonic data $\lambda^{1\omega} = 9.4 \pm 0.6$ μm. From the second harmonic signal (figure 3b,d), originating from magnons generated by heat produced in the injector strip, we find $\lambda^{2\omega} = 8.7 \pm 0.8$ μm. For distances larger than 40 μm, the non-local voltage is smaller than the noise level of our setup (approximately 3 nV$_{rms}$). We compare the magnitude and sign of the signal in the short distance measurements to a local measurement in Supplementary section A.

The first and second harmonic signal can be characterized by similar values of $\lambda$, indicating that thermally excited magnons are also generated in close vicinity of the injector. This supports the conclusions drawn by Giles et al.[13], i.e. for thermal magnon excitation the magnon signal reaches far beyond the thermal gradient generated by the applied heating. Note however that the sign change for the second harmonic signal (figure 3b inset) illustrates that the physics for electrical and thermal magnon generation is very different. This is discussed further in Supplementary section B.

We verify our assumption of magnon excitation and detection in linear response by performing measurements where we reversed the role of injector and detector. The results are shown in figure 4a (4b) for the first (second) harmonic. For the first harmonic non-local resistance we find $R_{V-I}^{1\omega} = 13.28 \pm 0.02$ m$\Omega$ and $R_{I-V}^{1\omega} = 13.26 \pm 0.03$ m$\Omega$. Since we find $R_{V-I}(B) = R_{I-V}(-B)$ [26], we conclude that Onsager reciprocity holds within the experimental uncertainty, despite the asymmetry in the injector-detector geometry. Reciprocity does not hold for the second harmonic (figure 4b), as expected for non-linear processes. Finally, we verify that $V_{nl}^{1\omega}$ scales linearly with applied current (figure 4c). The linearity and reciprocity of the first harmonic non-local signal demonstrate that it is due to linear processes only.

In conclusion, we report experimental proof for linear response magnon spin injection and detection in YIG via interaction with a spin accumulation in platinum. Additionally, we provide evidence for magnon transport in YIG over distances up to 40 μm, characterized by a magnon relaxation length $\lambda = 9.4 \pm 0.6$ μm at room temperature. Remarkably, the observed magnon transport is well described by the familiar spin diffusion model, despite the completely different character of the carriers of spins in magnetic insulators (bosons) compared to metals and semiconductors (fermions). Our results are consistent with spin injection/detection by invasive contacts, indicating that by optimizing contact properties the signals could be enhanced further.

## Methods

**Fabrication** The YIG samples used in the short distance device series (series A) consist of a 200 nm (111) single crystal YIG film grown on a 500 μm (111) $Gd_3Ga_5O_{12}$ (GGG) substrate by liquid-phase epitaxy (LPE), provided by the Université de Bretagne Occidentale in Brest, France. YIG samples used in the long distance series (series B) were obtained commercially from the company Matesy GmbH, and consist of a 210 nm single crystal (111) $Y_3Fe_5O_{12}$ film grown by LPE, also on a GGG substrate. The device pattern was defined using three e-beam lithography steps, each followed by a standard deposition and lift-off procedure. The first step produces a Ti/Au marker pattern, used to align the subsequent steps. The second step defines the platinum injector and detector strips, which were deposited by DC sputtering in an $Ar^+$ plasma at an argon pressure of $3.3 \times 10^{-3}$ mbar. The deposited Pt thickness was approximately 13 nm for series A devices and 7 nm for series B devices, measured by atomic force microscopy. The third step defines 5/75 nm Ti/Au leads and bonding pads, deposited by e-beam evaporation. Prior to Ti evaporation, argon ion milling was used to remove any polymer residues from the platinum strips, ensuring electrical contact between the platinum and the leads. Devices of series A have an injector/detector length of $L_A = 7.5$ to 12.5 μm and a strip width of $w_A \approx 100 - 150$ nm. Devices of series B have an injector/detector length of $L_B = 100$ μm and a strip width of $w_B \approx 300$ nm.

**Measurements** All measurements were carried out using three SR830 lock-in amplifiers using excitation frequencies ranging from 3 to 40 Hz. The lock-in amplifiers are set up to measure the first, second and third harmonic response of the sample. Current was sent to the sample using a custom built current source, galvanically isolated from the rest of the measurement equipment. Voltage measurements were made using a custom built pre-amplifier (gain $10^3$-$10^5$) and amplified further using the lock-in systems. The current applied to the sample ranged from 10 to 200 μA (root mean squared). The typical excitation current used is $I = 80$ μA, which results in a charge current density of $j_c \approx 10^{10}$ A/m², depending on the specific device geometry. The in-plane coercive field of the YIG is $B_c < 1$ mT for both YIG samples, and we apply an external field to orient the magnetization (typically $B_{ext} = 5$ mT) using a GMW electromagnet. The sample was rotated with respect to the magnet poles using a rotatable sample holder with stepper motor. The offset resistances $R_0^{1\omega}$ and $R_0^{2\omega}$ described in equations 1 and 2 depend on the capacitive and inductive coupling between the measurement wires to and from the sample and vanish for low excitation frequencies (typically when $f_{lock-in} < 5$ Hz).

## Acknowledgements

We would like to acknowledge M. de Roosz and J.G. Holstein for technical assistance, and would like to thank G.E.W. Bauer for discussions. This work is part of the research program of the Foundation for Fundamental Research on Matter (FOM) and supported by NanoLab NL, EU FP7 ICT Grant No. 612759 InSpin and the Zernike Institute for Advanced Materials.

## Author contributions

BJvW and LJC conceived the experiments. LJC designed and carried out the experiments, with help from JL. JBY supplied the YIG samples used in the fabrication of sample series A. LJC, JL, RAD and BJvW were involved in the analysis. LJC wrote the paper, with the help of the co-authors.

# References

1. Johnson, M. & Silsbee, R. H. Interfacial charge-spin coupling: Injection and detection of spin magnetization in metals. *Phys. Rev. Lett.* **55,** 1790–1793 (1985).

2. Lou, X. *et al.* Electrical Detection of Spin Transport in Lateral Ferromagnet-Semiconductor Devices. *Nat. Phys.* **3,** 197–202 (2007).

3. Tombros, N., Jozsa, C., Popinciuc, M., Jonkman, H. T. & van Wees, B. J. Electronic spin transport and spin precession in single graphene layers at room temperature. *Nature* **448,** 571–574 (2007).

4. Fabian, J., Matos-Abiague, A., Ertler, C., Stano, P. & Zutic, I. Semiconductor spintronics. *Acta Phys. Slovaca* **57,** (2007).

5. Zhang, S. S.-L. & Zhang, S. Magnon Mediated Electric Current Drag Across a Ferromagnetic Insulator Layer. *Phys. Rev. Lett.* **109,** 096603 (2012).

6. Zhang, S. S.-L. & Zhang, S. Spin convertance at magnetic interfaces. *Phys. Rev. B* **86,** 214424 (2012).

7. Uchida, K. *et al.* Spin Seebeck insulator. *Nat. Mater.* **9,** 894–7 (2010).

8. Kikkawa, T. *et al. Critical suppression of spin Seebeck effect by magnetic fields*. (2015). arXiv:1503.05764

9. Jin, H., Boona, S. R., Yang, Z., Myers, R. C. & Heremans, J. P. *The effect of the magnon dispersion on the longitudinal spin Seebeck effect in yttrium iron garnets (YIG)*. (2015). arXiv:1504.00895

10. Schreier, M. *et al.* Magnon, phonon, and electron temperature profiles and the spin Seebeck effect in magnetic insulator/normal metal hybrid structures. *Phys. Rev. B* **88,** 094410 (2013).

11. Schreier, M. *et al.* Current heating induced spin Seebeck effect. *Appl. Phys. Lett.* **103,** 2–6 (2013).

12. Kehlberger, A. *et al. Determination of the origin of the spin Seebeck effect - bulk vs. interface effects*. (2013). arXiv:1306.0784

13. Giles, B. L., Yang, Z., Jamison, J. & Myers, R. C. *Long range pure magnon spin diffusion observed in a non- local spin-Seebeck geometry*. (2015). arXiv:1504.02808

14. Kajiwara, Y. *et al.* Transmission of electrical signals by spin-wave interconversion in a magnetic insulator. *Nature* **464,** 262–6 (2010).

15. Slonczewski, J. C. Current-driven excitation of magnetic multilayers. *J. Magn. Magn. Mater.* **159,** L1–L7 (1996).

16. Brataas, A., Bauer, G. E. W. & Kelly, P. Non-collinear magnetoelectronics. *Phys. Rep.* **427,** 157–255 (2006).

17. Gurevich, A. & Melkov, G. *Magnetization oscillations and waves*. (CRC Press, 1996).

18. Serga, A. A., Chumak, A. V & Hillebrands, B. YIG magnonics. *J. Phys. D. Appl. Phys.* **43,** (2010).

19. Collet, M. *et al. Generation of coherent spin-wave modes in Yttrium Iron Garnet microdiscs by spin-orbit torque*. (2015). arXiv:1504:01512

20. Chumak, A. V. *et al.* Direct detection of magnon spin transport by the inverse spin Hall effect. *Appl. Phys. Lett.* **100,** 082405 (2012).

21. Bakker, F. L., Slachter, A., Adam, J.-P. & van Wees, B. J. Interplay of Peltier and Seebeck Effects in Nanoscale Nonlocal Spin Valves. *Phys. Rev. Lett.* **105,** 136601 (2010).

22. Valet, T. & Fert, A. Theory of the perpendicular mangetoresistance in magnetic multilayers. *Phys. Rev. B* **48,** 7099–7113 (1993).

23. Jungfleisch, M. B., Lauer, V., Neb, R., Chumak, A. V. & Hillebrands, B. Improvement of the yttrium iron garnet/platinum interface for spin pumping-based applications. *Appl. Phys. Lett.* **103,** 022411 (2013).

24. Vlietstra, N., Shan, J., Castel, V., van Wees, B. J. & Ben Youssef, J. Spin-Hall magnetoresistance in platinum on yttrium iron garnet: Dependence on platinum thickness and in-plane/out-of-plane magnetization. *Phys. Rev. B* **87,** 184421 (2013).

25. Takahashi, S. & Maekawa, S. Spin Injection and Detection in Magnetic Nanostructures. *Phys. Rev. B* **67,** 052409 (2003).

26. Onsager, L. Reciprocal Relations in Irreversible Processes. II. *Phys. Rev.* **38,** 2265 (1931).

# Figures

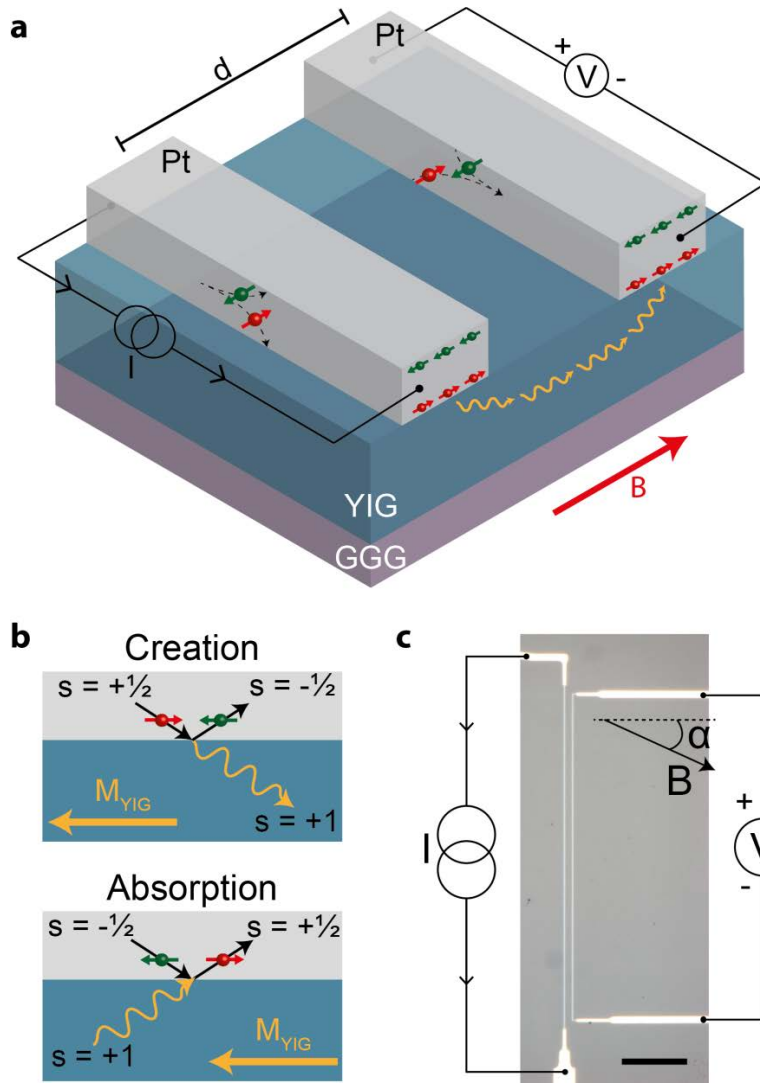

**Figure 1. Non-local measurement geometry. (a)** Schematic representation of the experimental geometry. A charge current *I* through the left platinum strip (the injector) generates a spin accumulation at the Pt|YIG interface through the spin Hall effect. Via the exchange interaction at the interface, angular momentum is transferred to the YIG, exciting or annihilating magnons. The magnons then diffuse towards the right platinum strip (the detector), where they are absorbed and a spin accumulation is generated. Via the inverse spin Hall effect the spin accumulation is converted to a charge voltage *V*, which is measured. **(b)** Schematic of the magnon creation and absorption process. A conduction electron in the platinum scattering off the Pt|YIG interface transfers spin angular momentum to the YIG. This will flip its spin and create a magnon. The reciprocal process occurs for magnon absorption. **(c)** Optical microscope image of a typical device. The parallel vertical lines are the platinum injector and detector, which are contacted by gold leads. Current and voltage connections are indicated schematically. An external magnetic field *B* is applied under an angle *α*. The scale bar represents 20 μm.

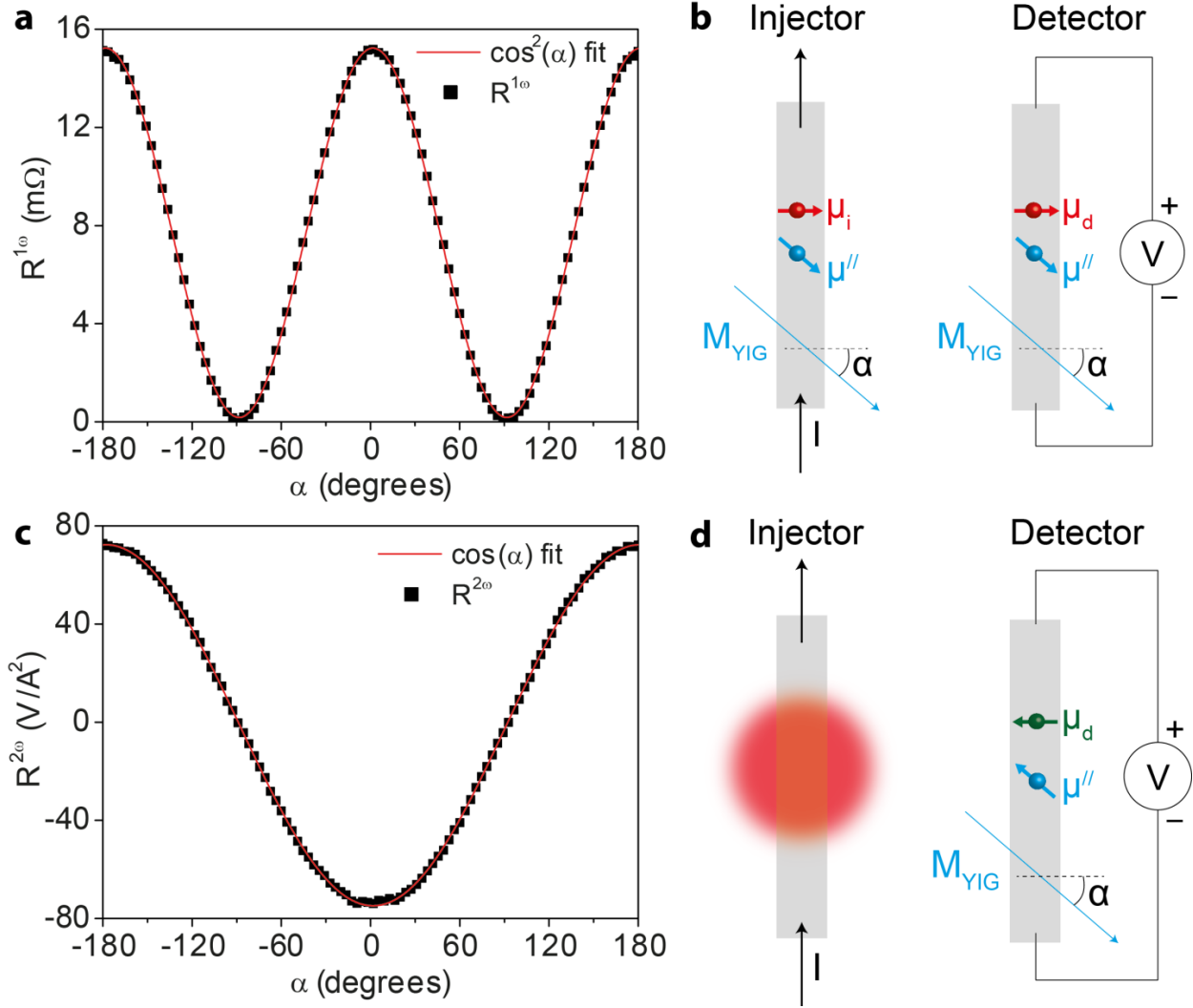

**Figure 2. Non-local resistance as a function of angle $\alpha$.** **(a)** First harmonic signal. The red line is a $\cos^2\alpha$ fit through the data. **(b)** Schematic top-view of the experiment. A charge current *I* through the injector generates a spin accumulation $\mu_i$ at the Pt|YIG interface. The component $\mu_\parallel$ parallel to the net YIG magnetization $M_{YIG}$ generates non-equilibrium magnons in the YIG, which gives rise to a $\cos\alpha$ dependence of the injected magnon density. The magnons then diffuse to the detector. At the detector, a spin accumulation $\mu_\parallel$ parallel to $M_{YIG}$ is generated. Due to the inverse spin Hall effect, $\mu_\parallel$ generates a charge voltage, of which we detect the component generated by $\mu_d$. This gives rise to a $\cos\alpha$ dependence of the detected magnon current. The total signal is a product of the effects at the injector and detector, leading to the $\cos^2\alpha$ dependence shown in figure a. **(c)** Second harmonic signal. The red line is a $\cos\alpha$ fit through the data. **(d)** Schematic illustration of the angular dependence of the second harmonic: Joule heating at the injector excites magnons thermally, which diffuse to the detector. This process is independent of $\alpha$. At the detector, the excited magnons generate a spin accumulation antiparallel to the YIG magnetization, which is detected in the same way as for the first harmonic, giving rise to a total $\cos\alpha$ dependence. The data shown in figure a and c is from a device with an injector-detector separation distance of 200 nm.

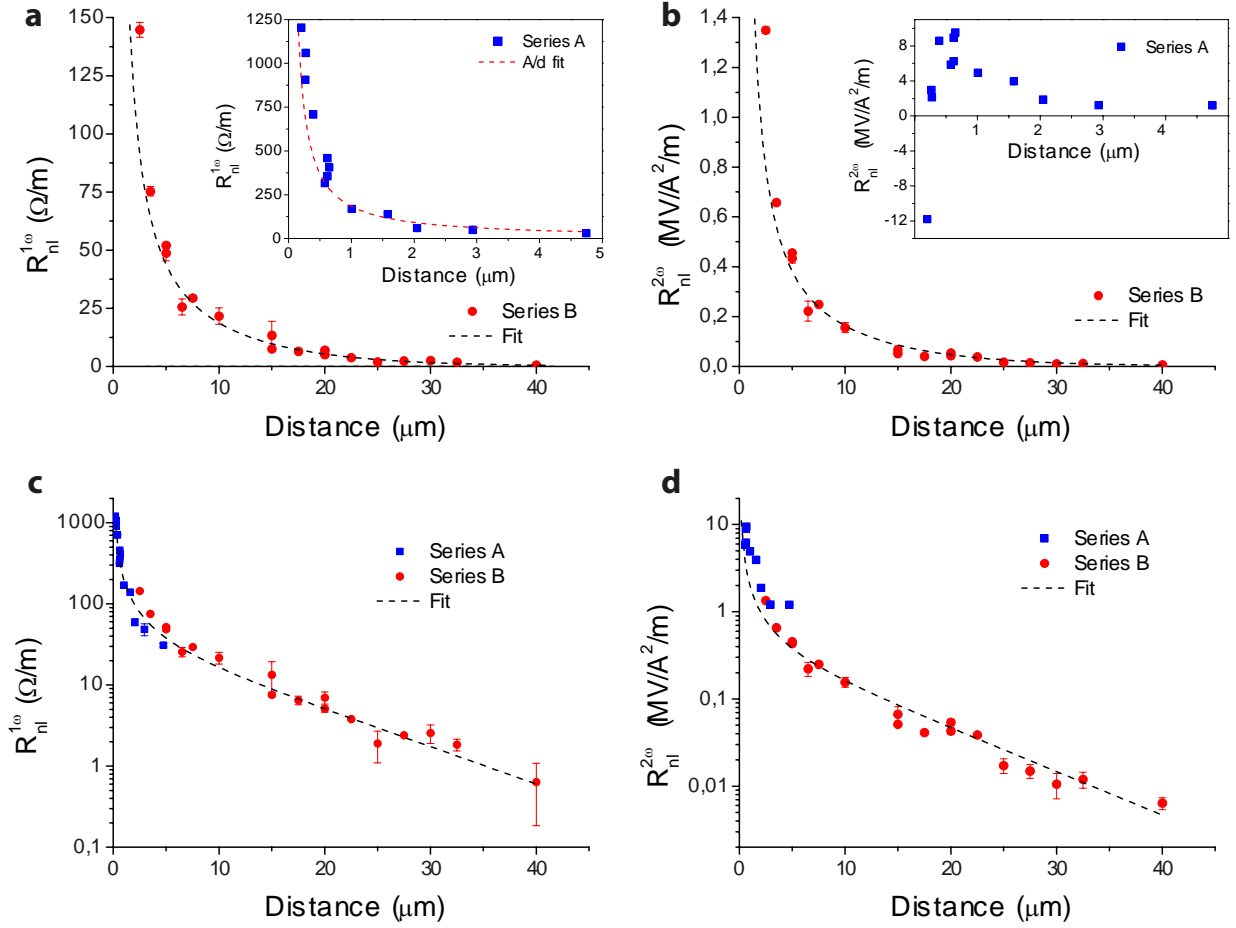

**Figure 3. Non-local resistance as a function of injector-detector separation distance.** Every data point represents a device with a different injector-detector separation and results from an angle-dependent measurement as shown in figure 2. The magnitude of the signal is extracted by fitting the angle-dependent non-local resistance to equation 1 (2) for the first (second) harmonic. The error bars represent the standard error in the fits. The signal is scaled by device length for both first and second harmonic. Figures **(a)** and **(c)** show the first harmonic data on a linear and logarithmic scale, respectively. The dashed line is a fit to equation 7, resulting in $\lambda^{1\omega} = 9.4 \pm 0.6$ μm. For $d < \lambda$, the data is well described by a $A/d$ fit, shown in the inset of figure a. Figures **(b)** and **(d)** show the second harmonic data on a linear and logarithmic scale. The dashed line is again a fit to equation 7, and we find $\lambda^{2\omega} = 8.7 \pm 0.8$ μm. The inset to figure b shows the short distance behavior of the second harmonic signal. For very short distances, the signal changes sign. For this reason, data points with $d < 0.5$ μm were omitted from the fit in figure b and d.

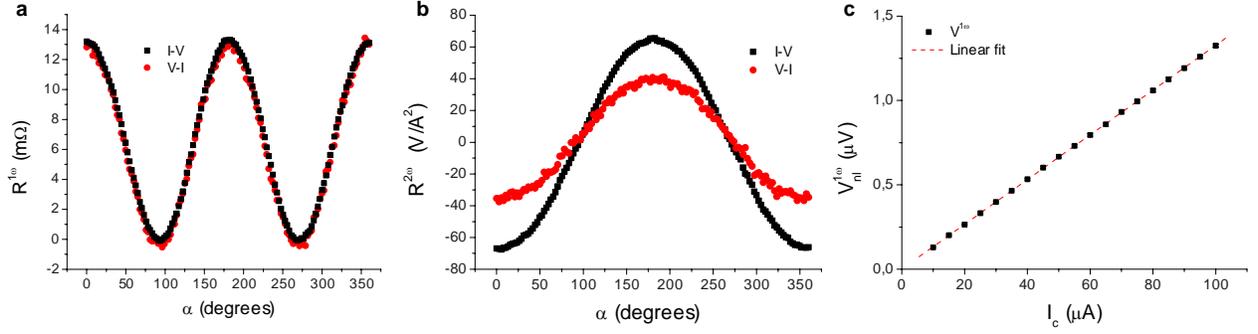

**Figure 4. Demonstration of reciprocity and linearity of the non-local resistance. (a)** First harmonic signal as a function of angle for the I-V and V-I configurations. We extract the amplitude of the signal using a fit to equation 1, and we find $R^{1\omega}_{V-I} = 13.28 \pm 0.024$ mΩ and $R^{1\omega}_{I-V} = 13.26 \pm 0.033$ mΩ. We thus conclude that reciprocity holds. **(b)** Second harmonic signal as a function of angle. We extract the amplitude of the signal using a fit to equation 2, and we find $R^{2\omega}_{V-I} = 76.3 \pm 0.5\ V/A^2$ and $R^{2\omega}_{I-V} = 132.8 \pm 0.5\ V/A^2$. We conclude that reciprocity does not hold for the second harmonic. **(c)** Non-local voltage as a function of injected charge current. The data is obtained from angle dependent measurements. A data point represents the amplitude of the angle dependent voltage, obtained by a fit to equation 1. The error bars, representing the standard error in the fit, are also plotted but are smaller than the data points. The red dashed line is a linear fit through the data, showing the linearity of the first harmonic signal.

# Supplementary Information

Long distance transport of magnon spin information in a magnetic insulator

## A. Comparison with local spin Hall magnetoresistance measurements

To study the limiting case $d \to 0$, we do a local spin Hall magnetoresistance (SMR) measurement[1] on the injector. In such a measurement, a current is sent through a platinum strip on a YIG film while simultaneously measuring the first harmonic voltage over that same strip. The resistance that is measured will depend on the orientation of the YIG magnetization **M** with respect to the spin orientation of the spin accumulation $\boldsymbol{\mu}_s$ in the platinum. When **M** $\perp$ $\boldsymbol{\mu}_s$, electron spins arriving at the Pt|YIG interface are absorbed. When **M** $\parallel$ $\boldsymbol{\mu}_s$, the spins are (mostly) reflected. This leads to a higher resistance for the perpendicular case than for the parallel case, and the difference in resistance $R^{SMR} = R_\perp - R_\parallel$ is the SMR response.

We do an SMR measurement on the injector strip of our devices. The sample is then rotated in an external field to extract the magnitude of the SMR response. We find $R_l^{SMR} = (6.8 \pm 0.2) \times 10^4$ Ω/m (averaged over 5 devices), approximately 57 times larger than the maximum non-local signal ($R_{nl}^{1\omega} = 1.2 \times 10^3$ Ω/m, for $d = 200$ nm). Note that with an SMR measurement we measure the difference between the number of absorbed spins and reflected spins, while in the non-local geometry we are sensitive only to the number of spins that are transferred across the interface when **M** $\parallel$ $\boldsymbol{\mu}_s$. Typically, the magnitude of the SMR signal is governed by the real part of the spin-mixing conductance $g_r$, while the number of transferred spins for the collinear case is governed by the effective mixing conductance[2] $g_s \approx 0.16 g_r$. For very short distances, the non-local signal will be dominated by the spin resistance of the Pt|YIG interfaces at the detector and injector (due to the finite $g_s$), and the ratio of the non-local short distance signal and the SMR

signal are in this case: $R_{nl}/R^{SMR} = g_s^2/g_r^2$. This gives $R^{SMR} = 39 \cdot R_{nl}$, which is close to what we observe. The fact that in our case $R^{SMR} \approx 57 \cdot R_{nl}$ shows that the non-local signal is reduced compared to the local signal further than expected. This is likely to be due to the spin resistance of the YIG channel, which can be non-negligible even for the very short channel length of 200 nm.

The symmetry of the SMR signal as a function of angle $\alpha$ is the same as that of the non-local resistance, since the SMR response also involves both the spin Hall effect and the inverse spin Hall effect, leading to the $\cos^2 \alpha$ dependence. Furthermore, the signs of the local and non-local signal agree, which is required since both effects involve the square of the spin Hall angle in platinum. Note that the sign of the non-local signal as described theoretically by Zhang & Zhang[3] is opposite from the sign we observe. This is still consistent with our observations however, since the parallel layer geometry (Pt|YIG|Pt) described in their paper yields an opposite direction for the spin current entering the platinum detector compared to our planar non-local geometry.

## B. Sign of the second harmonic response

As shown in the inset of figure 3b of the main text, the second harmonic signal changes sign for very short distances ($d = 200$ nm), while the signal is maximum for $d \approx 600$ nm. To investigate this sign change, we performed local measurements of the spin Seebeck voltage in the current heating configuration[4,5]. In this configuration, a charge current is applied to the injector and the second harmonic voltage over the injector is measured simultaneously. Figure S1 shows a comparison for the local (S1a), short distance non-local (S1c) and longer distance non-local (S1b, d) signals. It can be seen that the signal sign for the shortest distance matches that of the local spin Seebeck signal. We can understand this as when we move the detector closer and closer to the injector, we approach the limit of a local measurement. While it would be of great interest to develop a quantitative picture, this is outside the scope of this paper. Instead, we provide a qualitative explanation, guided by the fact that the change in sign occurs at a distance $d \approx 200$ nm which is comparable to the YIG film thickness $t_{YIG} = 200$ nm.

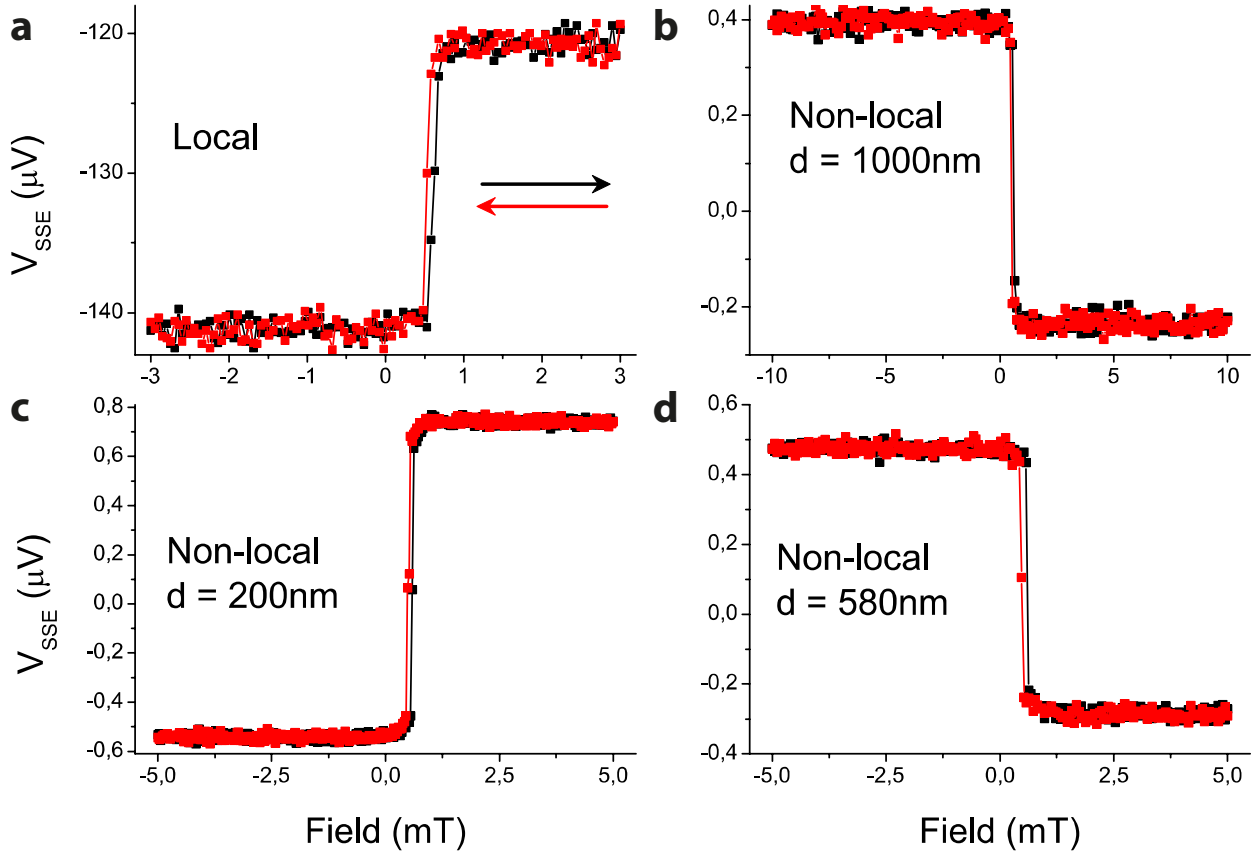

**Figure S1: Comparison of the sign of the second harmonic signal.** The spin Seebeck voltage $V_{SSE}$ is measured as a function of the external field, for a fixed sample position. Due to the $\cos \alpha$ symmetry of $V_{SSE}$, we observe a difference in signal for positive ($\alpha = 0$) and negative ($\alpha = -180°$) fields. From figure **a** and **c** it becomes clear that sign of the local $V_{SSE}$ agrees with that of the shortest non-local distance ($d = 200$ nm), while for longer distances (figures **b** and **d**) the sign is reversed. This can also be seen from the inset in figure 3b in the main text.

The second harmonic signal is due to thermally generated magnons, excited by the heat generated by Joule heating due to the current flowing in the injector strip. Since the heat conductivity of the GGG substrate is close to that of the YIG film, most of the heat generated in the injector will flow towards the substrate, normal to the plane of the film.

The spin Seebeck effect in Pt|YIG bilayers is well described as being driven by temperature differences at the Pt|YIG interface[6]. Additionally, a bulk spin Seebeck effect has been proposed where the driving force is the temperature gradient in the bulk of the YIG[7,8]. Following the bulk theory, an applied temperature gradient $\nabla T$ in a YIG film generates a magnon spin current parallel to $\nabla T$. Since the YIG film thickness is much smaller than the magnon spin relaxation length a magnon spin accumulation will build up at the interfaces, having opposite sign at the YIG|GGG compared to the Pt|YIG interface.

For the situation where $d$ is comparable to $t_{YIG}$, the magnon accumulation at the Pt|YIG interface dominates the signal, giving rise to an agreement in sign for the local and 200 nm non-local signal. For distances further away, diffusion of the YIG|GGG magnon accumulation across the film thickness compensates the Pt|YIG accumulation and becomes dominant for $d > t_{YIG}$, causing the sign reversal as a function of distance.

As a consequence, for distances $d < 500$ nm, the assumption that the injector is an ideal localized magnon source is no longer valid for the thermal generation case. Therefore, signal decay in this regime is no longer well described by equation 7 in the main text and we omitted the devices with $d < 500$ nm when fitting the second harmonic signal. Obtaining a quantitative picture of this behavior will require detailed modelling of the heat and spin currents in the sample.

# **References**

1. Nakayama, H. *et al.* Spin Hall Magnetoresistance Induced by a Nonequilibrium Proximity Effect. *Phys. Rev. Lett.* **110,** 206601 (2013).

2. Flipse, J. *et al.* Observation of the Spin Peltier Effect for Magnetic Insulators. *Phys. Rev. Lett.* **113,** 027601 (2014).

3. Zhang, S. S.-L. & Zhang, S. Spin convertance at magnetic interfaces. *Phys. Rev. B* **86,** 214424 (2012).

4. Vlietstra, N. *et al.* Simultaneous detection of the spin-Hall magnetoresistance and the spin-Seebeck effect in platinum and tantalum on yttrium iron garnet. **174436,** 1–8 (2014).

5. Schreier, M. *et al.* Current heating induced spin Seebeck effect. *Appl. Phys. Lett.* **103,** 2–6 (2013).

6. Xiao, J., Bauer, G. E. W., Uchida, K., Saitoh, E. & Maekawa, S. Theory of magnon-driven spin Seebeck effect. *Phys. Rev. B* **81,** 214418 (2010).

7. Rezende, S. M. *et al.* Magnon spin-current theory for the longitudinal spin-Seebeck effect. *Phys. Rev. B* **89,** 014416 (2014).

8. Hoffman, S., Sato, K. & Tserkovnyak, Y. Landau-Lifshitz theory of the longitudinal spin Seebeck effect. *Phys. Rev. B* **88,** 064408 (2013).